\documentclass[aps,twocolumn,prstper,showpacs,superscriptaddress]{revtex4-1}
\usepackage{graphicx}

\newcommand{\MCtwo}{Microtechnology and Nanoscience, MC2,
Chalmers University of Technology, SE-412 96 G{\"o}teborg, Sweden}
\newcommand{\Hule}{Huleb\"acksgymnasiet, 
Idrottsv\"agen 2, SE-435 80 M\"olnlycke, Sweden}
\newcommand{\norge}{Randaberg Videreg{\aa}ende Skole, Gr{\o}demveien 70,
NO-4070 Randaberg, Norway}

\begin{document}

\title{
Involving high school students in computational physics university research:\\
Theory calculations of toluene adsorbed on graphene}

\author{Jonas Ericsson}\affiliation{\Hule}
\author{Teodor Husmark}\affiliation{\Hule}
\author{Christoffer Mathiesen}\affiliation{\Hule}
\author{Benjamin Sepahvand}\affiliation{\Hule}
\author{{\O}yvind Borck}\affiliation{\norge}
\author{Linda Gunnarsson}\affiliation{\Hule}
\author{P\"ar Lydmark}\affiliation{\Hule}
\author{Elsebeth Schr{\"o}der}\thanks{Corresponding author}\email{schroder@chalmers.se}%
\affiliation{\MCtwo}

\date{March 28, 2016}
%%%%%%%%%%%%%%%%%%%%%%%%%%%%%%%%%%%%%%%%%%%%%%%%%%%%%%%%%%%%%%%%%%%%%%%%%%%%%%
\begin{abstract}
To increase public awareness of theoretical materials physics, 
a small group of high school students is invited to participate actively in a 
current research projects at Chalmers University of Technology.
The Chalmers research group explores methods for filtrating hazardous 
and otherwise unwanted molecules from
drinking water, for example by adsorption in active carbon filters.
In this project, the students use graphene as an idealized model for 
active carbon, and estimate the energy of adsorption of the methylbenzene toluene 
on graphene with the help of the atomic-scale calculational 
method density functional theory.
In this process the students develop an insight 
into applied quantum physics, a topic usually not taught at this 
educational level, and gain some experience with a couple of 
state-of-the-art calculational tools in materials research. 
\end{abstract}


\maketitle

%%%%%%%%%%%%%%%%%%%%%%%%%%%%%%%%%%%%%%%%%%%%%%%%%%%%%%%%%%%%%%%%%%%%%%%%%%%%%%

\section{Introduction}
In western societies, citizens are often not aware of research activities
at universities and other research institutions. This is a problem for a democratic society:  
research is often to some degree funded by the government, and 
results have bearings on how we live our lives, either applied directly or
in how we perceive the world. Lack of science awareness affects how we, as citizens, 
can discuss and make collective decisions.
 
It is important to convey research through education and outreach. 
One way to increase the public awareness is to let citizens 
participate in scientific research. 
School children and youth are natural targets for this as they are already  
engaged in 
learning. Their experience from research exposure or participation 
in younger ages will form their 
view on research also in later phases of their lives. 
Science centers and open-house days  
are ways to reach many people efficiently. 
Another way is to invite students to directly participate in research.  

We here report on computational physics research, 
in which a group of four senior-year high school students participated.

The students participated as part of a course at their school,
with a general introduction to research, given by their school, 
a one week visit to the research group, some months' work in class,
and a final report and poster session for classmates, parents, teachers, 
and research supervisors.

In the specific science project focused on here, the students carry out 
density functional theory (DFT) calculations 
to obtain the adsorption energy and structure of 
the toxic methylbenzene molecule toluene on graphene.

\begin{figure}
\begin{center}
\includegraphics[width=0.12\textwidth]{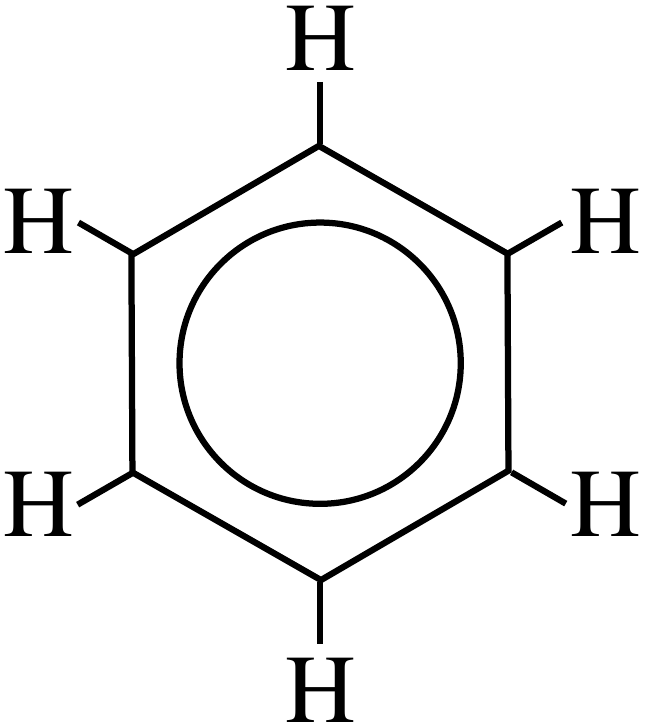}\hspace{0.1\textwidth}
\includegraphics[width=0.12\textwidth]{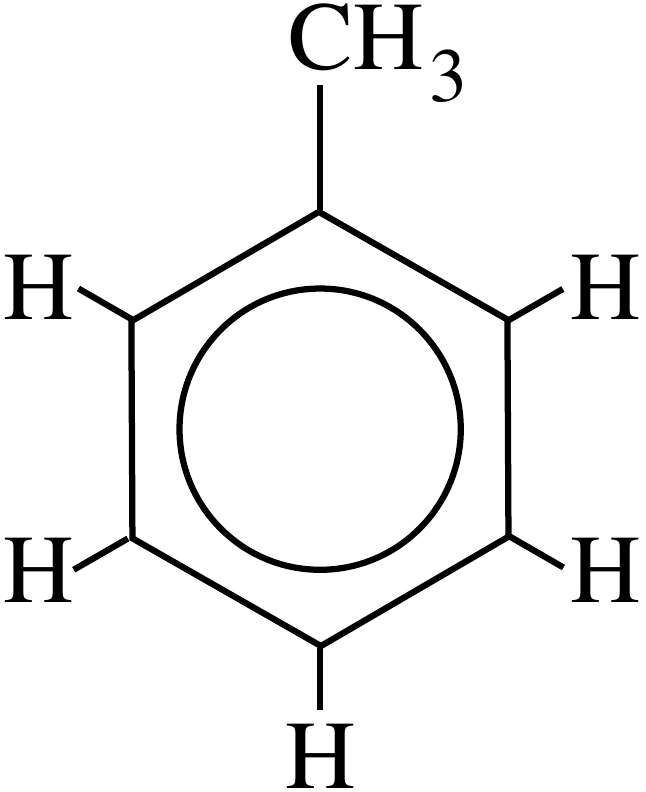}
\\[1em]
\includegraphics[width=0.12\textwidth]{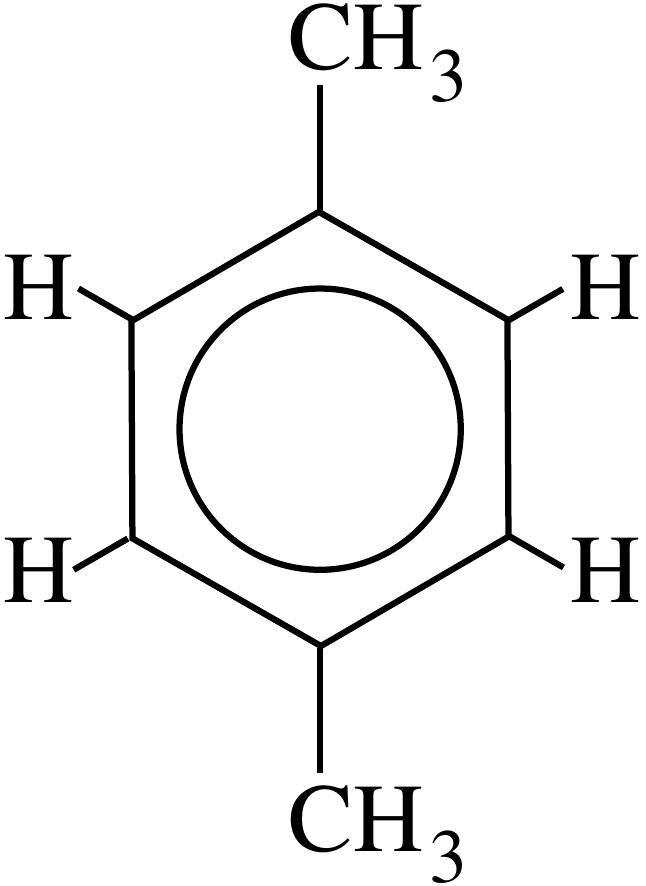}\hspace{0.1\textwidth}
\includegraphics[width=0.15\textwidth]{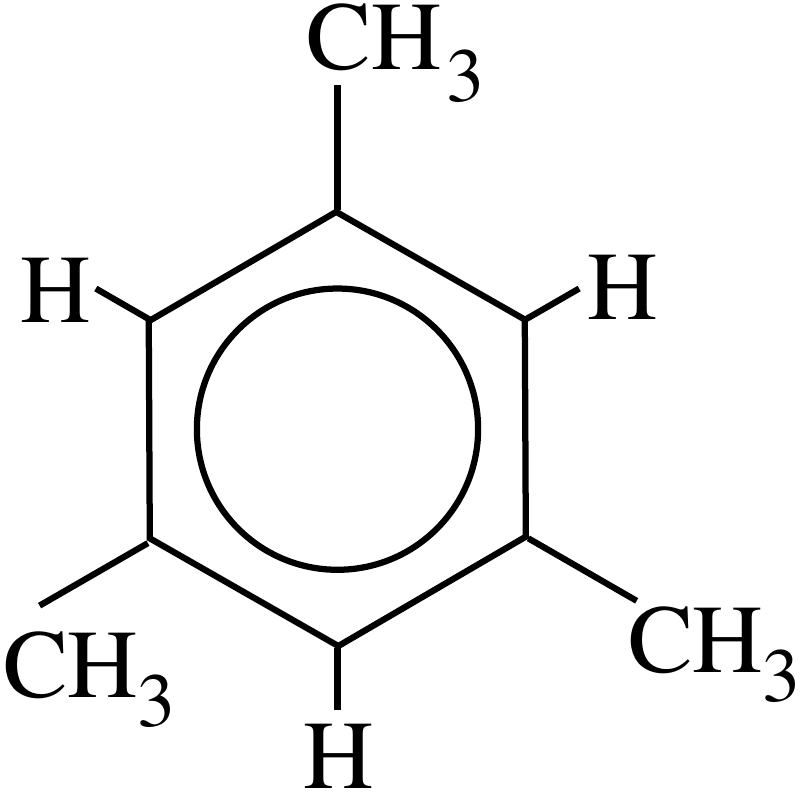}
\caption{\label{fig:btx}The atomic structures of benzene, and the 
methylbenzenes toluene, para-xylene, and mesithylene. 
The student research project focuses on toluene (top right), whereas the 
continuation of the project involves also benzene and para-xylene (bottom left), the
so-called BTX-family, as well as mesithylene (bottom right).}
\end{center}
\end{figure}

The choice of research topic is of course essential for successfully 
engaging the students. 
Understanding toxicity in our environment has proven one such focus that interests
our visitors. 
Methylbenzenes is a group of small, aromatic molecules that are volatile
and hazardous. They are benzene molecules with one or more methyl groups attached.
For the student collaboration we focus on toluene, a methylbenzene with one
methyl group, whereas a broader project of the Chalmers research group
 (reported elsewhere, Ref.~\onlinecite{borck15preprint})
includes also benzene and methylbenzenes with two and three methylgroups: 
para-xylene (1,4-dimethylbenzene)
and mesithylene (1,3,5-trimethylbenzene).
The atomic structures are shown in Figure \ref{fig:btx}.
 
In the Chalmers research group we study how carbon materials and other
filter materials may be used for filtering undesirable molecules from 
water or 
air.\cite{chakarova-kack06p146107,chakarova-kack06p155402,londero12p424212,schroder13p871706,akesson12p174702}
In our studies, smooth, defectless graphene is sometimes used as a first, idealized model of a filter,
and methylbenzene adsorption on graphene fits naturally into this program. 
For example, a similar study of chloroform adsorption on graphene was carried out by a preceding 
student group from the same high school program one year earlier, later extended
into a regular chemical physics publication.\cite{akesson12p174702}
The student part of the present computational project is trimmed and chosen such as to
minimize the computational time needed, and the results are used as a reference for
discussions of the necessary resources in such projects.

In the following we discuss in brief the overall research project of methylbenzenes
adsorbed on graphene, and then focus on the toluene student project and the implications
for the students and their education. 
In the appendix we describe the calculational 
limitations of the student project.
 
\section{The Science Problem}

Active carbon and similar materials are often used as filter materials for air and water filters.
 Although defects and impurities play an important role in 
how active carbon acts as an adsorbent, already the calculated adsorption energy for 
adsorption on clean, perfect graphene 
will be an indication of the strength of the adsorption in the filters.

Recently, the use of nanoporous graphene in reverse osmosis was suggested as a cost-effective
way to desalinate sea water.\cite{thiel15p66,surwade15p459}
The pressure needed during the reverse osmosis is lower with nanoporous graphene compared to
the use of conventional polymeric membranes, keeping down the energy consumption of the 
process.
The water molecules pass through the nanopores, leaving the larger salt 
ions behind in a growing concentration.
In this process pollutant molecules may stick to the graphene 
membrane,\cite{klomkliang12p5320} 
by which the water is then also cleaned for pollutants.

In the present project a number of methylbenzenes are used as examples of 
molecules that need to 
be filtered from water or air. Toluene, top right in Figure \ref{fig:btx}, is 
believed to be neurotoxic,\cite{white97p1239} and thus definitely not
suitable in drinking water. 

\section{Method of computation and results}

We here focus on the specific student project of toluene adsorption on graphene.
All calculations presented here 
use the calculational method
density functional theory\cite{kohn99} (DFT). It is a method for atomic-scale calculations
that builds
directly on the Schr\"odinger equation of quantum physics.
The input needed for the DFT calculations is in principle only the atomic numbers of
the atoms in the system and their approximate atomic positions, e.g., 
within the molecules or in the surfaces of the system.
In practical use a number of approximations are made, 
and the choice of approximations influence the size and duration of
the calculations as well as the quality of the results obtained.

In the present project, carried out by the high school students, 
the DFT method
vdW-DF\cite{dion04p246401,dion04p246401erratum,thonhauser07p125112,berland15p066501}  
in the version\cite{dion04p246401} vdW-DF1 is used,
as implemented in the code GPAW\cite{GPAWhttp,GPAW10} with ASE.\cite{ASEhttp,ASE02}
All calculations are carried out in computational clusters at C3SE, 
at Chalmers University of Technology, within the  
Swedish National Infrastructure for Computing (SNIC).

\begin{figure}
\begin{center}
\includegraphics[width=0.3\textwidth]{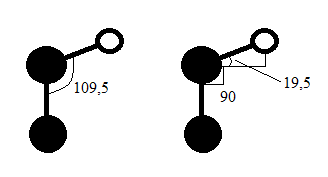}
\caption{\label{fig:angle}From the student report. Estimating the atomic positions
in toluene using symmetry considerations and the bond lengths of C-C and C-H, as well
as typical angles (in degrees) of the C-C-H bonds.
}
\end{center}
\end{figure}

Atomic positions for molecules are available from a range 
of sources online. Since the atomic structure of toluene is relatively
simple (Figure \ref{fig:btx}) the students are instead asked to 
estimate the atomic positions from knowledge of common C-C and C-H
bond lengths and relevant symmetries, Figure \ref{fig:angle}. The estimated 
atomic positions are then entered into 
a fast, but low-quality DFT calculation that moves the atoms such that 
the remaining net-forces on the atoms are minimized.
This changes the atomic positions slightly (and thus the bond lengths and
bond angles), and it is therefore not important to start out with 
precise values for the atomic positions.
For this part of the problem the computationally cheap 
linear-combination-of-atomic-orbitals mode (LCAO) is used. 
These atomic positions are then used as input for the further calculations,
with accurate grid-based wavefunctions.\cite{GPAWhttp}

\begin{figure}
\begin{center}
\includegraphics[width=0.4\textwidth]{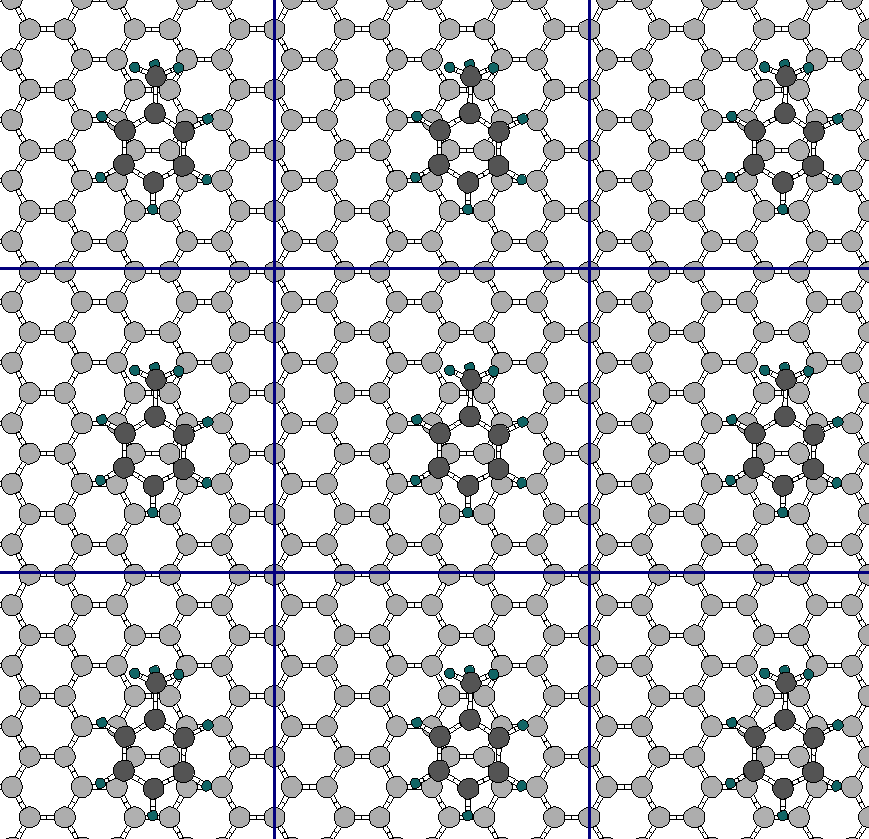}
\caption{\label{fig:repeat}Illustration 
of the repeated unit cell of the calculations, seen from an 
angle looking directly down on graphene.
The middle rectangle is the limit of one unit cell. The graphene 
C atoms are light gray large circles, the toluene
C atoms are dark gray large circles, and H atoms are small circles.  
The covalent bonds between atoms are a guide to the eye only, bonds 
are not explicitly treated in DFT calculations but occur 
naturally e.g.\ from the electron density distribution, as in nature.
The unit cell is orthorhombic (a box with right angles) with size 
$3\sqrt{3} a_g \times 5 a_g$  (12.9 {\AA} $\times$ 12.4 {\AA}) in the plane of graphene.
Here, $a_g=2.48$ {\AA} is the lattice parameter of the
primitive graphene unit cell in
our calculations. The size of the unit cell in the direction perpendicular to the 
plane of graphene is 23 {\AA}.}
\end{center}
\end{figure}

The system used for the toluene-on-graphene calculations includes 60 C atoms 
of graphene and the 15 atoms of toluene. 
However, in the calculations the unit cell is periodically repeated, so 
that the graphene is in effect infinite. The toluene molecule is
also periodically repeated, and thus there is a sparse network of toluene molecules at
fixed separation (equal to the side lengths of the unit cell) on graphene.
This is illustrated in Figure~\ref{fig:repeat}.

The toluene adsorption energy, $E_a$, is calculated as the difference in total system energy 
between the situation with the adsorbed molecule and the situation with toluene ``far''
away from graphene (in reality 11.5 {\AA} away)
\begin{equation}
E_a = -(E_{\mbox{\scriptsize bind}}-E_{\mbox{\scriptsize far}})
\label{eq:Ea}
\end{equation}
where positive values are obtained if toluene binds.

\begin{figure}[tb]
\begin{center}
\includegraphics[width=0.4\textwidth]{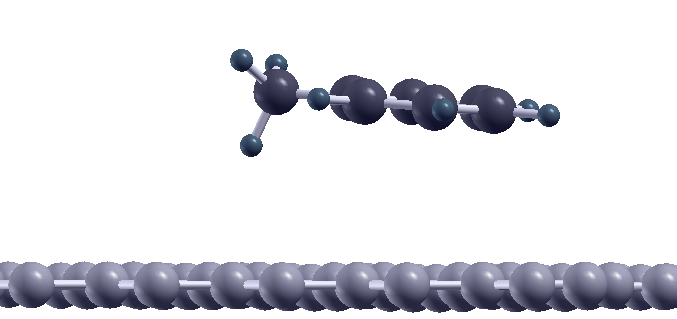}
\caption{\label{fig:side}Illustration 
of the toluene molecule adsorbed on graphene, seen perpendicular 
to the plane of graphene. Here the configuration found by the students
is shown, with one H atom of the methyl group pointing towards graphene,
and the other two methyl-group H atoms pointing away (termed ``methyl corner'' in the text
and table). 
}
\end{center}
\end{figure}

In Appendix \ref{app:A} we describe the simplifications used in the 
student project in order to shorten the student calculations,
so that they fill only the available time (one week at the university campus).  
Together, all simplifications reduce week-long
calculations to calculations of 10--30 minutes, albeit at the loss of some of 
the accuracy. As will be discussed further below, the accuracy is still sufficient for the
students to obtain meaningful results.  

\begin{table}[tb]
\caption{\label{tab:energies}Adsorption energies $E_a$ for
toluene on graphene calculated with various values of parameters,
of real-space grid points (gpts), number of Brillouin zone k-points (kpts),
and exponent of energy convergence threshold $n$, $1.5\cdot 10^{-n}$ eV/electron.
The student research project provided the
``student'' data for toluene, whereas the medium- and high-quality toluene
data were obtained later by the research group.\protect\cite{borck15preprint} As a comparison,
results of experimental measurements from the 
literature\cite{ulbricht06p2931,monkenbusch81p442} are also shown,
however, these results cannot be directly compared, as discussed in the main text.
}

\begin{tabular}{lcccc}
 \hline\hline
     & gpts  & kpts &  $n$  &$E_a$ [eV] \\
  \hline
\textit{methyl corner} & \\
  student     & $60 \times 60 \times 96$  & $1\times 1\times 1$ & 6 & 0.48 \\ 
  medium qual.& $108\times 100\times 192$ & $2\times 2\times 1$ & 6 & 0.50 \\  
  high quality  & $108\times 100\times 192$ & $4\times 4\times 1$ & 7 & 0.49 \\ 
\textit{methyl edge} & \\
  medium qual.& $108\times 100\times 192$ & $2\times 2\times 1$ & 6 & 0.52 \\ 
  \hline
  exper., Ref.~\protect\onlinecite{ulbricht06p2931}   & & &  & $0.71\pm 0.07$ \\
  exper., Ref.~\protect\onlinecite{monkenbusch81p442}&&&  & $0.52 $ \\
\hline
\end{tabular}
\end{table}

The adsorption energies $E_a$ are calculated from eq.~(\ref{eq:Ea}) and are shown in
Table \ref{tab:energies}.  
We see that for the particular configuration (one methyl H atom pointing down, ``methyl corner'') 
the $E_a$ of the student calculations deviates from the more 
accurate, but also more costly calculations of the same configuration,
by 0.01--0.02 eV (1--2 kJ/mol), or less than 4\%. While the accuracy of the
student results is not quite publication quality, the effect on the binding
energy is marginal, and shows us that such short calculations can indeed 
be used in student projects and still give realistic results. 

The students used a configuration where one of the methyl-group H atoms points
towards graphene and the other two point away. The triangle formed
by the three H atoms in the methyl group thus points one of its corners
towards graphene. The configuration is therefore termed ``\textit{methyl corner}''
in Table~\ref{tab:energies}, and is illustrated in Figure~\ref{fig:side}. 
In the subsequent work of the research group\cite{borck15preprint}
configurations with instead two of the methyl-group H atoms pointing
towards graphene and the third pointing away (here termed ``\textit{methyl edge}'', 
because an edge of the H atom triangle points to graphene)
was found to be slightly more favourable for the
adsorption energy. However, even compared to the \textit{methyl edge} configuration
the student calculations differ in energy by less than 6\% (Table~\ref{tab:energies}). 

\begin{figure}[tb]
\begin{center}
\includegraphics[width=0.45\textwidth]{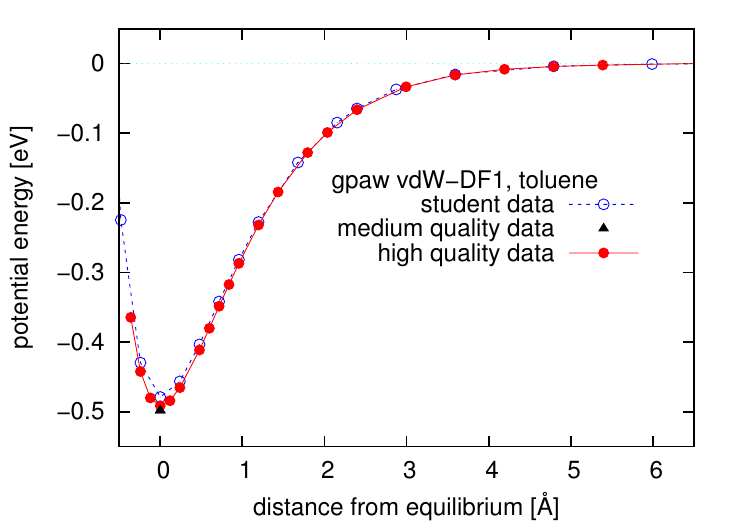}
\caption{\label{fig:PES}Potential energy curve for toluene flat on top of 
graphene at various separations.
Curves for high quality data as well as the student-produced data with 
lower quality on the calculational parameters, as discussed in the main text.
All data are for the configuration with one methylgroup H atom pointing towards
graphene (``methyl corner'').}
\end{center}
\end{figure}

To refine this discussion, we also include the 
 interaction energy (potential energy curve)
of toluene above graphene at various distances, shown in Figure \ref{fig:PES}.
The students carried out the calculations that are marked with open dots, 
while the highest 
quality calculations are shown with filled dots.
An intermediate-quality calculation, improving only partly on the student 
settings, is shown as a triangle, and interestingly this intermediate
calculation shows results slightly away from both the student results and the
high-quality results. This illustrates the fact that the path to convergence
in parameter setting is not necessarily monotonous, and for full convergence 
care must be taken. 
In the adsorption situation toluene is slightly tilted with respect to 
the graphene plane (as seen in Figure~\ref{fig:side}), 
thus in Figure \ref{fig:PES} we use the
molecule center of mass for indicating the distance to graphene.

Experimental measurements\cite{ulbricht06p2931,monkenbusch81p442} 
of the heat of adsorption are also included in Table \ref{tab:energies}.
The experimental results cannot be directly compared to the calculated results
for a number of reasons, but they do indicate that the calculated results are reasonable.
Contrary to the calculations, the experiments report on results from 
 systems that are not in vacuum, are at non-zero temperature, and are more densely packed
at the surface, with a close to
one monolayer (ML) coverage. The 1 ML coverage in experiment is to be contrasted to the 
approximately 0.24--0.29 ML coverage in our
calculations,\footnote{Ref.~\onlinecite{monkenbusch81p442} reports that the area occupied by an
adsorbed toluene molecule on graphite is measured to be 46 {\AA}$^2$, 
of which we have one per 12.9 {\AA} $\times$ 12.4 {\AA} area,
leading to a coverage of $46/160 \approx 0.29$ ML. Ref.~\onlinecite{klomkliang12p5320} 
finds from molecular dynamics calculations that the coverage concentration is
4.30 $\mu$mol/m$^2$, which means that the area per molecule is 38.6 {\AA}$^2$ and 
which would mean that our coverage is 0.24 ML.} 
that are further treated such as
to obtain the values for a single molecule on graphene (Appendix A).
We know from other similar systems\cite{schroder13p871706} that the adsorption energy increases
with coverage.

Our results do not include the zero-point (quantum mechanical) fluctuations, 
i.e., the quantum mechanical vibrations that are present even at zero temperature. 
Based on a parabola fit to the potential energy curve in Figure~\ref{fig:PES} 
we estimate the zero-point motion perpendicular to
graphene to contribute a few meV to the total zero-point motion of toluene on graphene. 
Assuming that the lateral parts contribute similar amounts, the change in calculated adsorption energies
is barely noticeable. We therefore ignore the zero-point fluctuations here.

Finally, 
another difference to the experiments is that the experiments are adsorption 
on graphite, which is a stack of sheets of graphene, and not on a single layer, 
graphene.
The effect from the lower-lying layers in graphite is thus not included 
in our graphene calculations.
It has previously been estimated that adsorption on graphite leads to 
roughly 3\% higher interaction energy compared to the same adsorption in 
graphene.\cite{chakarova-kack06p146107}
Altogether, the 8--30\% difference between experiments and theory is reasonable.

\section{Project context and preconditions}

With an increasing pressure on university faculty to publish more, 
obtain more external funding, and teach more efficiently, it is no wonder that reaching out to society 
may seem like an additional burden in a tightly scheduled  
work day. Nevertheless, outreach is important, and can often be formed such that it ends
up benefiting also the faculty member, in addition to the public. Reaching out to 
school children and other youth, like high school students and undergraduates, has a 
special value because they already are in a learning process.

Research with undergraduates is well established in
the science, technology, engineering, and mathematics (STEM) field. 
Involving school children and high school youth, on the other hand, 
is less common, and the barriers are considered significant.
In fact, involving high school youth in the STEM field seems to
pose similar barriers to those for 
undergraduate research in the humanities, eloquently described by 
History professor C.\ R.\ Corley in Ref.~\onlinecite{corley13p397}.
Including these groups of students in research work is seen as a use of 
time that is not really available. 
Exceptions exist, for example large-scale simple experiments like 
the international Tea Bag Index\cite{keuskamp13p1070} used with 
Swedish school children, the best-before-date food study in the fridges of 
Swedish homes by school children,\cite{marklinder15p1764} and 
similar initiatives through the
Researchers Night.\cite{forskarfredag}
In these projects a large number (hundreds) of pupils contribute to real research by
adding their data points, like the degree of tea leave degradation after months embedded in soil,
or the temperatures of the family fridge at various positions and the condition of the food. 

In literature on undergraduate research the importance of being able to contribute to 
original research is emphasized. In Ref.~\onlinecite{osborn09} the main characteristics of undergraduate
research are listed as mentorship, originality, acceptability, and dissemination. 
For research with younger students, it is natural to expect the conditions 
to be fulfilled to a lesser degree, but encompassing all four elements 
makes for a real research experience.
 
The student project presented here is a project for four high school students, as 
part of a course on research, given by the high school (Huleb\"acksgymnasiet).
The course involves a general introduction to science research, measurements in
class for a small example project (bouncing ball),  a week in a research group 
on university campus with members of the research group,
and a possibility of further contact with the faculty member 
before and after the week-long stay, if needed.
Finally, the students write a report and organize a poster presentation.

The four main characteristics of undergraduate research, mentioned by 
Osborn and Karukstis,\cite{osborn09} are
to varying degree also covered by the type of high school research described here: 
Prior to the week on campus, the research project is introduced (short visit
or email) and relevant reading material is suggested. 
During the first day of the one-week visit, focus is on helping the students navigate 
the computer codes, understand (to some degree) how they work, and getting a very
short introduction to the theory behind the project, in this case quantum physics and
density functional theory. As the days go by, the students are expected to 
work increasingly independent.  
Thus the \textit{mentoring\/} has the characteristics of teaching in the beginning
of the week, but crosses over to mentoring of more independent nature towards 
the end of the week. 
Typically, the students will have the possibility to choose direction of the project after
the initial results are in. In the present project the students chose to calculate the
full potential-energy curve presented in Figure \ref{fig:PES}, after having obtained the
adsorption energy and successfully having translated between the various units involved, 
from J/mol to eV, and jumping between length units of {\AA}ngstr\"om and Bohr, 
in order to compare results to the literature. 

Making sure that the research is \textit{original\/} is the responsibility of the faculty member,
and should not be of consideration for the students until they reach a point where they
suggest further calculations beyond the initial results. 
The research project in itself may be an \textit{acceptable\/} project 
from the research community point of
view, but care needs to be taken that the quality of the results is also acceptable. 
This may pose problems with the one-week time frame, but as discussed in Appendix A,
in the present project ``clever'' choices of parameters can lower the time 
needed without significant loss of accuracy, and results can be verified 
later with higher accuracy by the faculty participants.

The \textit{dissemination\/} of the students' results is mainly via the two student
products: the school report and the poster. In some cases the results are also
published in scientific journals, after further work by the faculty mentors.\cite{akesson12p174702}

The students involved in these projects do this as an alternative way
of carrying out their high school diploma project, offered as a special 
course called Forskarkursen (``The science course'') at the high school 
Huleb{\"a}cksgymnasiet. Yearly 16--24 students are given the 
opportunity to apply to the course.
Students are not selected based on their grades, but it is a  
prerequisite to be highly motivated and to have the ability to 
grasp the complexness of science projects.

The course starts at the high school a few months prior to
the university project week with one hour of lesson per week. 
Here the students are informed about the academic system, 
trained to read scientific papers and discuss ethic scientific dilemmas. 
In addition, the students  perform a small, open investigation answering 
the question ``How many times does a ball bounce?'' to illustrate the scientific 
method when faced with a problem that has no given answer.

After the project week at the university the students analyze
their data and write their scientific report with guidance of the 
high school teachers and the university supervisor. 
The course has been given for more than ten years, and the course evaluations have 
always been very positive. The students especially appreciate that they get to 
perform real research that is of immediate interest for a research group. 
They are also encouraged by the fact that they actually manage to understand 
the essence of the specific research field they are working in, even though 
it is well above their high school science level. 
Examples of such research projects, 
besides the one described in this paper, are
 how shorthorn sculpin react to temperature changes 
(Department of Zoology, University of Gothenburg), 
yeast cell responses to oxidative stress 
(Department of Chemistry and Molecular Biology, University of Gothenburg), 
gene expression of enterotoxigenic Escherichia coli
(Department of Microbiology and Immunology, Sahlgrenska Academy, University of Gothenburg), 
and visualization of 1P and 2P fluorescence
(Department of Physics, University of Gothenburg).

The one-week research-group visit focuses on carrying out the research for the project,
 but time is also spent seeing the working place and feeling part of the work environment.
The roles of the teachers and mentors are relatively clear: the faculty mentor
works with the students on campus, whereas the high school teachers work
with the students on general areas in school. The high school teachers involved
in this course both have a background as researchers in the STEM field,
which benefits the interaction between the high school teachers and the 
university mentors.

The overall content of the project must be carefully chosen.
One must be certain to obtain usable results (i.e., negative results 
should still be of interest, in order to not discourage the students), 
thus research that measures a number is a better choice than testing 
a hypothesis that is only interesting if positive.
It must be something new, but the group should still be able to easily 
compare at least some of the results to previous work/experiments
in the area.

For the above reason, the present project is similar to previous projects that have
been carried out in the research group: it studies the adsorption of small, flat molecules onto
a flat surface. Previous such research results, using the same or similar
computational tools, include calculations
of benzene, adenine, phenol, and the other nucleobases on 
graphene.\cite{chakarova-kack06p146107,berland11p135001,chakarova-kack06p155402,le12p424210}
From a calculational point of view toluene has the
advantage of being a rather small molecule with a relatively stiff part in the
aromatic ring. This keeps the need for calculational power and time down,
compared to studies of more complicated adsorbants.

On the practical, computational side of the project,
the DFT code (GPAW\cite{GPAWhttp,GPAW10}) 
and the python computational environment (ASE\cite{ASEhttp,ASE02})
are open-source codes and as such carry the benefit that the students at any time
can download the codes and play around with them at their own computers. A national computational cluster is
used for parallelized computations in order to obtain results in a reasonable time, but smaller
such calculations can also be carried out at the laptops of the students, given
sufficient patience. The fact that this is possible (even if not actually downloaded and installed)
helps to demystify the research process.

\section{Summary}
We present an example of organized research for high school students, given
as a course at the Swedish high school Huleb{\"a}cksgymnasiet, in collaboration
with local universities. A specific research project within computational physics 
is discussed both from the outreach and the scientific point of view.

\acknowledgments
Support from the Swedish Research Council (VR) and from
the Knut and Alice Wallenberg Foundation (KAW) is gratefully acknowledged.
The computations were in part performed on resources at Chalmers Centre for
Computational Science and Engineering (C3SE) provided by the
Swedish National Infrastructure for Computing (SNIC).
The student research course is a collaboration between
Huleb\"acksgymnasiet, G\"oteborg University, and Chalmers
University of Technology, with 
high school course teachers and supervisors
LG and PL and with
project adviser ES for this particular group of 
students.
Researcher and high school teacher {\O}B joined ES in 
obtaining the medium- and high-quality results, and the (then) high school
students JE, TH, CM and BS obtained the ``student data'' as part of their
research experience.

\appendix

\section{Simplifications in the student calculations \label{app:A}}

The adsorption system is described within an orthorhombic unit that 
is periodically repeated in all directions, Fig.~\ref{fig:repeat}.
Each toluene molecule is therefore affected by the presence of the
neighboring toluene molecules. However, the calculations of binding energy are carried out so as to cancel
most of this possible contribution to the energy, yielding the binding energy of a single 
toluene molecule on graphene. 

The DFT density of the electrons (and the corresponding quantum-mechanical wavefunctions) are
described on a uniform grid. The distance between grid points in the student project
is almost double that in the research project.
Thus the student project only has 17\% of the grid points compared to the research projects.

One further difference is the use of only one k-point (the $\Gamma$ point) in the student
calculations whereas for the research calculations we use $2\times 2 \times 1$ 
or $4\times 4\times 1$ k-points.
The k-points are the discretization of the reciprocal space, used for easy evaluation of
a double gradient in the kinetic energy.
The use of only the $\Gamma$ point is common, and correct, in calculations of
material systems that are not extended, e.g.,
in the interaction of two small molecules. Here one part of the system is confined (toluene)
and the other part is extended (graphene). Thus it is not unreasonable to use the
$\Gamma$ point only, although for accuracy research calculations usually do make use 
of a more dense grid of k-points. By use of time reversal symmetry $4\times 4\times 1$ k-points can be
reduced to 4 points in the irreducible Brillouin zone, four times more than just the
$\Gamma$ point.  

Since the computational need grows faster than linearly with number of grid points and 
number of irreducible k-points the student calculations can be carried out with 
(much) less than 4\% (1/4 of 17\%) of the resources required for the high-quality
calculations. This makes otherwise week-long calculations on the computational cluster
reduce to few-hour calculations on less computations nodes, making the project attainable
for a one-week student project. 

Finally, the high-quality calculations differ from the student (and medium-quality calculation)
by a stricter convergence condition on the energy calculated: the calculation is considered 
converged if changes are less than $1.5\cdot 10^{-n}$ eV/electron, where the requirement is
$n=7$ for the high-quality but $n=6$ for the medium-quality and student calculations.
This does change the computational time needed, but much less than the other sources above,
and this difference will be ignored here. 

\bibliography{methods_vdWpapers.bib,paperspecific.bib}

\end{document}